\documentclass[superscriptaddress, amsmath, amssymb, aps, twocolumn]{revtex4-2}
\usepackage{graphicx}
\usepackage{bm}
\usepackage{color}
\usepackage{siunitx}
\usepackage{mathtools}
\usepackage{ulem}
\usepackage{float}
\usepackage{hyperref}
\usepackage[all]{hypcap}
\hypersetup{colorlinks,allcolors=blue}

\begin{document}

\title{High-Efficiency spin-Seebeck Diode in an $\alpha'$-Borophene FM/Normal/FM Nanoribbon Junction}

\author{Farzaneh Ghasemzadeh}
\affiliation{\footnotesize{Department of Physics, Iran University of Science and Technology, Narmak, Tehran 16844, Iran}}

\author{Mohsen Farokhnezhad}
\email{m.farokhnezhad@saadi.shirazu.ac.ir}
\affiliation{\footnotesize{Department of Physics, College of Sciences, Shiraz University, Shiraz 71946-84795, Iran}}

\author{Mahdi Esmaeilzadeh}
\email{mahdi@iust.ac.ir}
\affiliation{\footnotesize{Department of Physics, Iran University of Science and Technology, Narmak, Tehran 16844, Iran}}

\date{\today}
\begin{abstract}
The $\alpha'$-borophene nanoribbon ($\alpha'$-BNR) due to its incredible properties such as high stability and great mobility of carriers demonstrates high-efficiency in thermoelectric devices. We show that these properties enable us to produce a pure spin current by applying a temperature gradient with lower energy consumption in a ferromagnetic/normal/ferromagnetic (FM/Normal/FM) junction. Spin-dependent thermoelectric properties and spin-Seebeck are studied in this junction using the tight-binding (TB) formalism in combination with the non-equilibrium Green's function method (NEGF). The pure spin current due to the breaking of the electron-hole symmetry is induced in the system so that it can act as a spin-Seebeck diode. Moreover, the negative differential spin-Seebeck effect can be observed in the system. Finally, we show, at the same conditions, the $\alpha'$-BNR has a much higher power factor compared to that of graphene and silicene, which is due the high asymmetry between the electrons and holes in the $\alpha'$-BNR. The exceptional features of $\alpha'$-BNR makes it a very suitable choice for using in thermoelectric devices.\end{abstract}

\maketitle
\section{Introduction}
The emerging field of spin caloritronics \cite{elahi2023recent,yang2023role,uchida2021spintronic,hirohata2020review,dey2021spintronics} which combines spintronics with thermoelectronics, has attracted considerable interest due to its potential for enhancing energy conversion efficiency~\cite{pawar2023recent,vzutic2004spintronics,wolf2001spintronics, mishra2025borophene}. Thermal-spin devices are capable of producing a spin current under a temperature gradient, even without applying an external electric field. The control of spin currents can be achieved through the regulation of thermal currents. Under certain circumstances, the spin Seebeck effect can generate a pure spin current within selective materials by breaking of the electron-hole spin symmetry \cite{wu2021spin,bader2010spintronics}. The observed phenomenon depends on the spin-dependent Seebeck coefficients, which exhibit contrasting polarities for spin-up and spin-down electrons. Such an intrinsic property of materials plays a pivotal role in the development of nanodevices that demonstrate exceptional efficiency and longevity with low power consumption. The low-dimensional nanostructures such as two-dimensional (2D) materials and nanoribbons have received noticeable attention due to their low phonon thermal conductivity and the high spin-dependent Seebeck coefficients, which result in higher thermoelectric efficiency compared to bulk materials~\cite{hicks1993effect,dresselhaus1999advances,heerema2018probing}.

Borophene, a planar structure comprised of boron atoms~\cite{piazza2014planar, norouzi2021controllable}, has attracted considerable attention among researchers owing to its diverse arrangements and notable functionalities including electronic properties, optical transparency, and high electron mobility~\cite{li2018review,hou2023borophene,sergeeva2014understanding,mannix2018borophene,hou2020borophene,nasir2019emerging,kaneti2021borophene, ghasemzadeh2024ultrafast}. Several successful experimental syntheses of borophene have been reported on metallic substrates such as Ag, Cu, Ni, and Au \cite{wu2019large, kiraly2019borophene, zhong2017synthesis}. This material is being acknowledged as a potential competitor to graphene. Furthermore, the distinct band gap shown by different phases of borophene imparts advantageous practical properties in contrast to the gapless nature of graphene~\cite{sergeeva2014understanding,mannix2018borophene,zhang2017two,wu2012two}. Upon comparing graphene with the 8B-Pmmn phase of borophene, it becomes evident that the latter has a superior carrier mobility of $10^6$ cm$^2$V$^{-1}$ s$^{-1}$ at room temperature conditions, which is more than one order of magnitude greater than that of graphene (the carrier mobility of graphene is 15000 cm$^2$V$^{-1}$ s$^{-1}$)~\cite{cheng2017anisotropic,chen2022recent}. 

The formation of delocalized multicenter (nc-2e) bonds in boron, arising from its electron-deficient nature has led to the formation of a wide variety of borophene phases \cite{zhong2016towards, norouzi2021controllable}. Recently, three experimentally borophene phases ($\beta_{12}$, $\chi_{3}$ and $\alpha'$) have been successfully synthesized on MoS$_2$ substrates \cite{hou2021ultrasensitive}. Interestingly, first-principle calculations have confirmed that these phases retain their pristine electronic properties on MoS$_2$ \cite{kang2024selective}. Although most borophene phases (e.g., $\beta_{12}$, $\chi_{3}$ and $\alpha$) exhibit highly metallic behavior due to high electron density near the Fermi level \cite{ranjan2020borophene, zhang2021semiconducting}, the first-principles calculations have confirmed that the $\alpha'$-phase of borophene sheet is semiconducting, with an indirect band gap of approximately 0.25 eV \cite{zhang2021semiconducting}. Using the first-principles calculations, Wu \textit{et al.} reported that $\alpha$-phase of borophene is dynamically unstable, as indicated by the presence of imaginary frequencies in the out-of-plane acoustic (ZA) phonon branch \cite{wu2012two}. This instability suggests that the planar $\alpha$ structure tends to reconstruct into a slightly buckled $\alpha'$-phase, which is more stable due to lower total energy compared to the $\alpha$-phase \cite{zhang2021semiconducting}. This buckled structure of borophene with hollow hexagons in its configuration due to the high cohesive energy, high mobility of electrons, and semiconducting behavior makes the $\alpha'$-borophene sheets an important 2D material in nanotechnology~\cite{wu2012two,zhang2021semiconducting,tang2007novel,yang2008ab,penev2012polymorphism}. The high stability of $\alpha'$-borophene is due to the weak buckling which is made of the hybridization between $\pi$-bond and $s+p_{x,y}$ orbitals \cite{zhang2021semiconducting}.

Using the first-principles calculations, the thermal transport property of the $\alpha'$-borophene sheet was studied by Xiao \textit{et al.}. They reported that the lattice thermal conductivity of the $\alpha'$-borophene sheet is about 14.34 Wm$^{-1}$K$^{-1}$ which is much smaller than that of graphene, which is about 3500 Wm$^{-1}$K$^{-1}$~\cite{xiao2017lattice}. This low thermal conductivity of the $\alpha'$ sheet supports a higher figure of merit than that of graphene. Zhang \textit{et al.}, investigated the excellent stability of $\alpha'$-borophene sheet using a tight-binding (TB) model combined with first-principles calculations \cite{zhang2021semiconducting}. Their results illustrate, due to the small effective mass of electrons and holes, the carrier mobility in the $\alpha'$-borophene sheet is high enough which can be an appealing candidate for manufacturing 2D field effect transistors (FETs). Low Ohmic contact resistance between the semiconducting $\alpha'$-borophene channel and the metallic borophene as the electrodes is another properties of this material~\cite{zhang2021semiconducting}.

The 1D structures of materials, known as nanoribbons, have demonstrated superior properties in the fields of electronic and thermoelectric transport. In recent years, extensive research have been made to exploring the thermoelectric properties of various nanoribbons~\cite{rodrigues2022exploring,xie2023intrinsic,rahmati2022thermoelectric,li2023spin,tan2020thermal,ildarabadi2021spin}. Metals were the first materials investigated for thermoelectric applications. Despite their high electrical conductivity, they generally exhibit low Seebeck coefficient values and high thermal conductivity, rendering them inappropriate for efficient thermoelectric conversion. In contrast, semiconductors owing to large magnitude of Seebeck coefficients are usually the ideal candidates for thermoelectric devices, which arise from the presence of a band gap that breaks the electron-hole symmetry. This feature suggests that materials with semiconducting behavior can achieve high thermoelectric efficiency with a large power factor (PF). Moreover, the high carrier mobility enhances the thermoelectric performance due to weaker electron-electron and electron-phonon interactions~\cite{markov2019thermoelectric}. 

The thermoelectric properties of the semiconducting $\alpha'$-borophene nanoribbons ($\alpha'$-BNR) have not yet been systematically investigated. Thus, we intend to study the thermopower and spin-Seebeck effect in $\alpha'$-BNRs in more detail. In this paper, we have designed a thermoelectric device composed of ferromagnetic (FM) electrodes and a normal channel. The calculation of spin-dependent currents and the spin thermopower are performed via the non-equilibrium Green's function (NEGF) method and the Landauer-B\"uttiker formula. Our results show that $\alpha'$-BNRs can generate a pure spin current under a temperature gradient due to strong electron-hole asymmetry. The proposed FM/Normal/FM device exhibits spin-Seebeck diode behavior as well as a negative differential spin-Seebeck effect. In addition, although both charge and spin power factors decrease with increasing temperature, the $\alpha'$-BNR still shows superior thermoelectric performance compared with graphene and silicene nanoribbons. These findings suggest that $\alpha'$-BNRs are promising candidates for thermoelectric and spin-caloritronic applications.    

The article is arranged as follows: in Sec. II we present our system and model which are employed to calculate the thermoelectric properties. The results of this study are summarized in Sec. III. Finally, we conclude our results and findings in Sec. IV.  

\begin{figure}[h]
\centering
\includegraphics[scale=0.27]{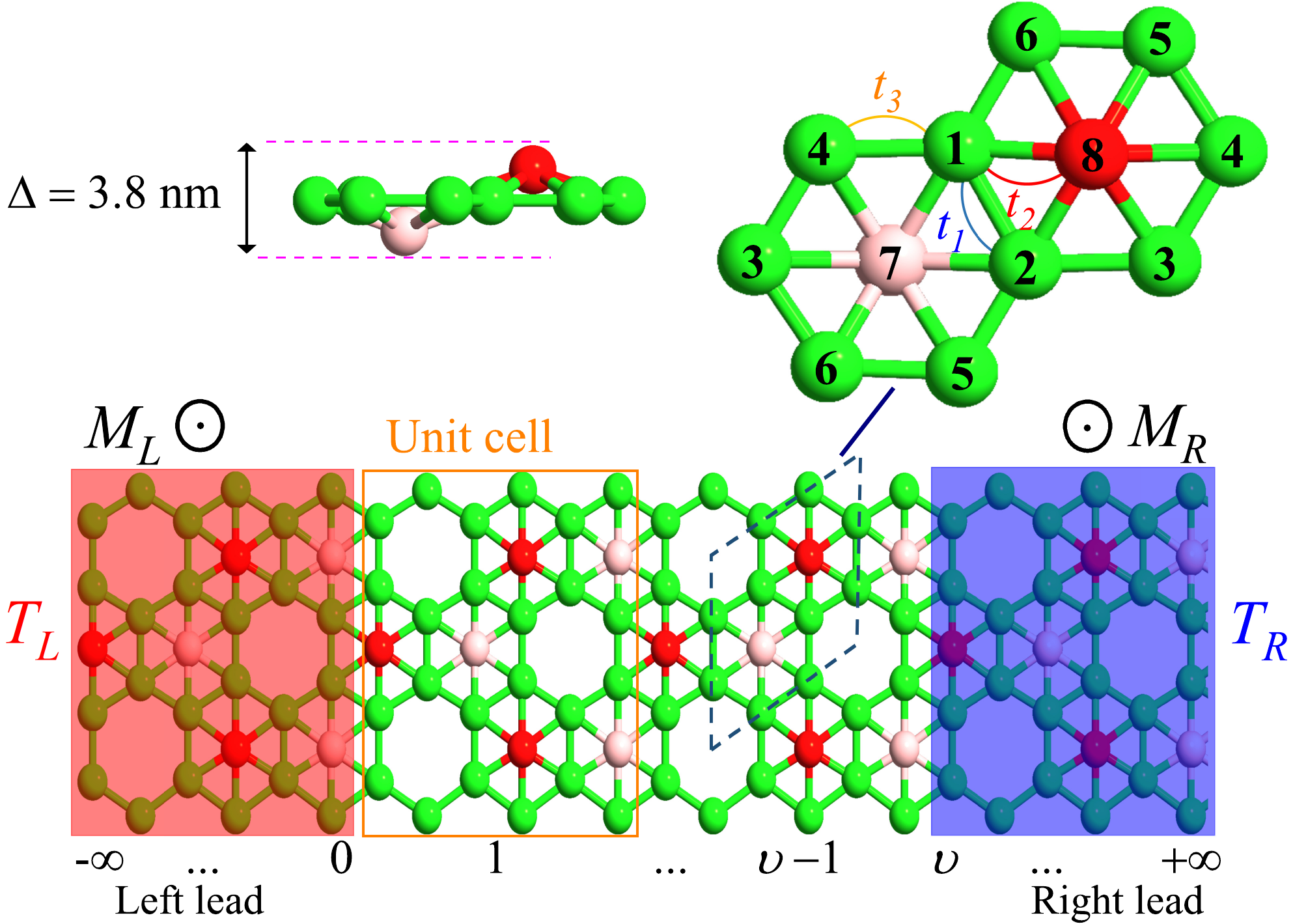}
\caption{\label{Fig1} Schematic depicting of a zigzag $\alpha'$-BNR-based FM/Normal/FM junction, consisting of a scattering part (channel) and the same two semi-infinite ferromagnetic (FM) leads. The magnetic moment of the two FM leads is aligned at the z-axis ($M_R=M_L=M_z$) and $T_L$ and $T_R$ represent the temperatures of the left and right leads, respectively. The primitive cell is indicated by the blue dashed lines. Nearest-neighbor (NN) hopping energies are represented by $t_1$, $t_2$ and $t_3$. The channel contains $\nu$ unit cells, each consisting of a total of \textit{N} boron atoms. Here, the number of atoms in each unit cell and the strength of exchange field of FM leads are $N=30$ and \textit{$M_z$}=0.098 eV, respectively.}
\end{figure}

\section{THEORETICAL MODEL} \label{sec2}

We consider a zigzag $\alpha'$-BNR-based  FM/Normal/FM junction, consisting of a central channel connected to two semi-infinite leads as shown in Fig.~\ref{Fig1}. Magnetization in the leads is induced by the proximity effect arising from a ferromagnetic insulator (such as EuO \cite{swartz2012integration}) deposited on their surface. The magnetization orientations in both electrodes are assumed to be identical and aligned along the z-axis (i.e., $M_R=M_L=M_z$). The total Hamiltonian of the system, $H_T$, can be written as
\begin{equation}\label{Eq.1}\tag{1}
H_T=H_C+H_L+H_R,
\end{equation}
where $H_C, H_L$ and $H_R$ are the Hamiltonians of the scattering region (channel), the left and right FM leads, respectively. The TB Hamiltonian for each part of the system (channel, left and right electrode) is expressed as
\begin{equation}\label{Eq.2}\tag{2}
H_C=-\sum_{\langle{i,j}\rangle,\alpha}{t_{ij}}c_{i\alpha}^{\dagger}c_{j\alpha}+{\sum_{{\langle{i,j}\rangle},\alpha}}{\varepsilon_{i}}c^{\dagger}_{i\alpha}c_{j\alpha},
\end{equation}
\begin{eqnarray}
H_L=H_R=-\sum_{\langle{i,j}\rangle,\alpha}{t_{ij}}c_{i\alpha}^{\dagger}c_{j\alpha}+{\sum_{{\langle{i,j}\rangle},\alpha}}{\varepsilon_{i}}c^{\dagger}_{i\alpha}c_{j\alpha}\nonumber\\
+M_z\sum_{i,\alpha}c^{\dagger}_{i\alpha}\sigma_{z}c_{i\alpha},
\label{Eq.3}
\end{eqnarray}
where $t_{ij}$ in the first term denotes the hopping energy between the nearest-neighbor (NN) sites. The notation ${\langle{i,j}\rangle}$ represents the sumation over NN hopping pairs. The operators $c_{i\alpha}^{\dagger}$ and $c_{i\alpha}$ are the creation and annihilation operators, respectively, which create and annihilate an electron with spin $\alpha$ at site $i$. The hopping parameters \textit{$t_1$} (between the two atoms of the adjacent sides of the hexagonal, atom 1 to 2), \textit{$t_2$}(between the central atom with each corner atom, atom 1 to 8), and \textit{$t_3$} (between the two atoms on the hollow hexagon's sides, atom 1 to 4) are shown in Fig.~\ref{Fig1}. The second term denotes the on-site energy ${\varepsilon_{i}}$. There are two different types of on-site energy where ${\varepsilon_{1}}$ and ${\varepsilon_{2}}$ are the on-site energy of fivefold (five-bonding) (1 and 2) and sixfold(six-bonding) atoms (7 and 8), respectively. The hopping and on-site energy parameters, calculated by fitting to first-principles results, are summarized as ~\cite{zhang2021semiconducting}\\[8pt]
 $  t_{1}=2.55$ eV,  \:          $ t_{2} = 2.05$ eV, \:     $ t_{3}= 2.19$ eV, \:     \\[8pt]   	
 $\varepsilon_{1} = -1.90$ eV,  \:\: $ \varepsilon_{2} = -2.65$ eV. \hfill(4)        \\[8pt]    
The final term in Eq.~\ref{Eq.3} describes an exchange field with magnitude $M_z$, generated by the proximity effect of ferromagnetic materials on the leads~\cite{bishnoi2013spin}. Moreover, $\sigma_{z}$ is the z-component of the Pauli matrices. Notice that spin-orbit interactions are generally weak in light-element materials; therefore, they are negligible in boron-based systems.

The spin-dependent conductance can be calculated using the NEGF method. For this purpose, the surface retarted Green's functions of the left and right leads are obtained through the use of an iterative technique (i.e., known as Lopez-Sancho's algorithm) as \cite{sancho1984quick,li2008quantum}\\[8pt]
$\text{g}^{L}_{0,0}(E)=[(E+i\eta)\textbf{I}-H_{0,0}-H^{\dagger}_{-1,0}\widetilde{T}]^{-1}, $\hfill(5)\\[8pt]
$\text{g}^{R}_{\nu,\nu}(E)=[(E+i\eta)\textbf{I}-H_{\nu,\nu}-H_{\nu,\nu+1}{T}]^{-1}, $\hfill(6)\\[8pt]
where \textbf{I} is an identity matrix, $\eta$ is an infinitesimal positive real number, $H_{0,0}$ $(H_{\nu,\nu})$ is the Hamiltonian matrix of the assumed unit cell at site 0 $(\nu)$ of the device, and $H_{-1,0}$ $(H_{\nu,\nu+1})$ is the coupling matrix of the left-hand (right-hand) adjacent cells at sites -1 and 0 ($\nu$ and $\nu+1$). In this study, the left lead has been considered the same as the right one; $H_{0,0}=H_{\nu,\nu}$ and $H_{-1,0}=H_{\nu,\nu+1}$. The transfer matrices \textit{T} and $\widetilde{T}$ are determined through the following sequences~\cite{sancho1984quick,li2008quantum,nardelli1999electronic}:\\[8pt]
$T=\tilde{t}_{0}+\tilde{t}_{0}{t}_{1}+\tilde{t}_{0}\tilde{t}_{1}{t}_{2}+\ldots+\tilde{t}_{0}\tilde{t}_{1}\tilde{t}_{2}\ldots {t}_{n}, $\hfill(7)\\[8pt]
$ \widetilde{T}=t_{0}+{t}_{0}\tilde{t}_{1}+{t}_{0}{t}_{1}\tilde{t}_{2}+\ldots+{t}_{0}{t}_{1}{t}_{2} \ldots \tilde{t}_{n}, $\hfill(8)\\[8pt]
where the definition of $ t_{i} $ and $\tilde{ t_{i}}$ through the recursion relations can be written as\\[8pt]
$ t_{i}=(\textbf{I}-t_{i-1}\tilde{t}_{i-1}-\tilde{t}_{i-1}t_{i-1})^{-1}t_{i-1}^{2}, $ \hfill(9)\\[8pt]
$\tilde{t_{i}}=(\textbf{I}-t_{i-1}\tilde{t}_{i-1}-\tilde{t}_{i-1}t_{i-1})^{-1}{\tilde{t}_{i-1}^{2}}, $ \hfill(10)\\[8pt]
and\\[8pt]
$ t_{0}=[(E+i\eta)\textbf{I}-H_{00}]^{-1}H_{-10}^{\dagger},$\hfill(11)\\[8pt]
$\tilde{t_{0}}=[(E+i\eta)\textbf{I}-H_{00}]^{-1}H_{-10}.$\hfill(12)\\[8pt]
The above iteration procedure is repeated until ${t}_{n}$ and $\tilde{t}_{n}$ have a tendency to a quite small arbitrary amount.

Now, the surface Green's function inside the transport channel can be obtained, step by step, through the sampling of the scattering region as a part of the right lead returning from $l=\nu$ to $l=2$, using the recursion relation as follows~\cite{li2008quantum}:
\begin{equation}\tag{13}
\text{g}^{R\alpha}_{l,l}(E)=[(E+i\eta)\textbf{I}-H_{l,l}-H_{l,l+1}\text{g}^{R\alpha}_{l+1,l+1}
H^{\dagger}_{l,l+1}]^{-1}.
\end{equation}\\
Furthermore, the total ${\text{g}_{11}^{\alpha}}$ Green's function can be written as~\cite{datta2005quantum}
\begin{equation}\tag{14}
\text{g}_{11}^{\alpha}=[(E+i\eta)\textbf{I}-H_{11}-\Sigma_{L}^{\alpha}-\Sigma_{R}^{\alpha}]^{-1}.     \:\:\:   
\end{equation}
The self-energy of the left $(\Sigma_{L}^{\alpha})$ and the right $(\Sigma_{R}^{\alpha})$ leads  are defined by
\begin{equation}\tag{15}
\Sigma_{L}^{\alpha}=H_{01}^{\dagger}\text{g}_{00}^{L\alpha}H_{01},
\end{equation}
\begin{equation}\tag{16}
\Sigma_{R}^{\alpha}=H_{12}\text{g}_{22}^{R\alpha}H^{\dagger}_{12}.
\end{equation}
The density of states can be calculated using the Green's function as follows:
 \begin{equation}\tag{17}
\rho(E)=-\frac{1}{\pi}\text{Im}[\text{Tr}(\text{g}_{11}(E))].
\end{equation}
\\ Finally, the spin-dependent conductance can be witten by using the the Landauer-B\"uttiker formula as~\cite{datta2005quantum, farokhnezhad2015controllable} 
\begin{equation}\tag{18}
G^{\alpha}(E)=\frac{e^2}{h}T^{\alpha}(E),     \:\:\:     \alpha=\uparrow,\downarrow.
\end{equation}
where $T^{\alpha}(E)$ is the spin-dependent transmission coefficient and is given by \cite{fisher1981relation,meir1992landauer}
\begin{equation}\tag{19}
T^{\alpha}(E)=\text{Tr}[\Gamma_{L}^{\alpha}(E){\text{g}^{\alpha}_{11}}(E)\Gamma_{R}^{\alpha}(E)(\text{g}^{\alpha}_{11}(E))^{\dagger}].
\end{equation}
The broadening matrix which represents the coupling interaction between the channel and the left (right) lead can be written as 
\begin{equation}\tag{20}
\Gamma_{L(R)}^{\alpha}=i\left({\Sigma_{L(R)}^{\alpha}-(\Sigma_{L(R)}^{\alpha})^{\dagger}}\right).
\end{equation}
The spin-dependent current induced by the temperature gradient $\Delta{T}=T_L-T_R$ between the left and right leads is described by the Landauer-B\"uttiker formula~\cite{sivan1986multichannel}:
\begin{equation}\tag{21}
I^{\alpha}=\frac{e}{h}\int_{-\infty}^{+\infty}T^{\alpha}(E)[f_L(E,T_L)-f_R(E,T_R)] dE. 
\end{equation}
The spin current ($I_S = I^{\uparrow}-I^{\downarrow}$) and charge current ($I_C = I^{\uparrow}+I^{\downarrow}$) are defined as the difference and the sum of the spin-up $I^{\uparrow}$ and spin-down $I^{\downarrow}$ currents, respectively. Further, $T^{\alpha}(E)$ denotes the spin-dependent transmission coefficient, while $f_L$ and $f_R$ are the Fermi-Dirac distribution functions of the left and right leads at temperatures $T_L$ and $T_R$, respectively. The Fermi-Dirac distribution function at temprature \textit{T} and energy \textit{E} is given by
\begin{equation}\tag{22}
f(E,T)=\frac{1}{1+e^{(E-E_F)/k_BT}}, 
\end{equation}
where $E_F$ and $k_B$ are the Fermi energy and Boltzmann constant, respectively. The spin-Seebeck thermopower coefficient in the linear regime (i.e., $\Delta{T}\ll{T_L}$) is given by~\cite{sivan1986multichannel}
\begin{equation}\label{Eq.23}\tag{23}
S^{\alpha}(E_F,T)=-\frac{1}{|e|T}\frac{L_{1\alpha}(E_F,T)}{L_{0\alpha}(E_F,T)}, 
\end{equation}
where the spin-dependent function $L_{n\alpha}(E_F,T)$ is expressed as~\cite{sivan1986multichannel}
\begin{equation}\tag{24}
L_{n\alpha}(E_F,T)=-\frac{1}{h}\int_{-\infty}^{+\infty}T^{\alpha}(E)\frac{\partial f(E,T)}{\partial E} dE, 
\end{equation}
and the derivative of the Fermi-Dirac distribution function with respect to energy is
\begin{equation}\tag{25}
\frac{\partial f(E,T)}{\partial E}=-\frac{1}{k_BT}f(E,T)(1-f(E,T)).
\end{equation}
The spin and the charge thermopower coefficients can be written as~\cite{rameshti2015spin}
\begin{equation}\tag{26}
\text{S}_{s}=S^{\uparrow}-S^{\downarrow},
\end{equation}
\begin{equation}\tag{27}
\text{S}_{c}=\frac{(S^{\uparrow}+S^{\downarrow})}{2}.
\end{equation}
Finally, the important property known as the spin (charge) power factor in thermoelectric is given by
\begin{equation}\tag{28}
PF_{s(c)} = S_{s(c)}^2\sigma_{s(c)}, 
\end{equation}
where $\sigma_{s(c)}$ is the spin (charge) conductivity of the system and can be obtained as follows ~\cite{rameshti2015spin}:
\begin{equation}\tag{29}
\sigma_s =|\sigma^{\uparrow}-\sigma^{\downarrow}|,
\end{equation}
\begin{equation}\tag{30}
\sigma_c =\sigma^{\uparrow}+\sigma^{\downarrow},
\end{equation}
and the spin-dependent conductivity is expressed as~\cite{ghasemzadeh2024thermal}
\begin{equation}\tag{31}
\sigma^{\alpha} =e^2L_{0\alpha}.
\end{equation}
\section{RESULTS AND DISCUSSION}\label{sec3}
\begin{figure}[h]
\centering
\includegraphics[scale=0.48]{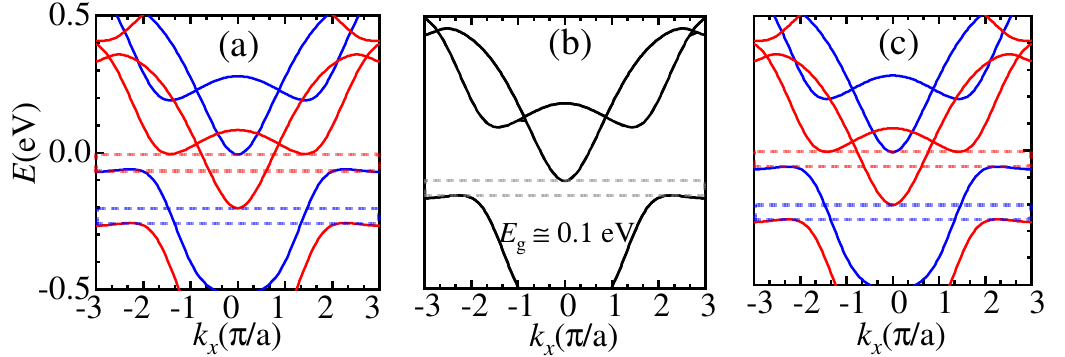}
\caption{\label{Fig2} Band structures of the (a) left lead, (b) channel, and (c) right lead for a zigzag $\alpha'$-BNR 
junction with \textit{N} = 30 in the presence of a magnetization $M_z$ = 0.098 eV applied in a parallel configuration to both leads. The blue and red solid lines correspond to the spin-up and spin-down bands, respectively. In Fig.~\ref{Fig2}(b), the spin-up and spin-down bands are degenerate, as indicated by the black solid lines. The regions indicated by blue and red dashed lines correspond to energy windows with perfect spin polarization for spin-up and spin-down states, respectively.}
\end{figure}

The results of numerical calculations based on TB Hamiltonian and the NEGF formalism are now presented in this section. Choosing that the electron mean free path in borophene structures is about 36 nm~\cite{norouzi2021controllable} and considering a channel length of $L=20.5$ nm, ensures ballistic transport in our proposed system (see Fig.~\ref{Fig1}). The right and left leads are magnetized by the ferromagnetic insulator substrates arranged in a parallel configuration, inducing the exchange magnetic fields of strength $M_z$ = 0.098 eV in both leads, alingned along the $z$-axis. In this case, the band structure of the left lead, channel region and the right lead are shown in Figs.~\ref{Fig2}(a)-(c). As illustrated in Figs.~\ref{Fig2}(a) and~\ref{Fig2}(c), due to the breaking time-reversal symmetry (TRS) through the use of an exchange magnetic field, the spin-up and spin-down bands are split and move in opposite directions. The channel region lacks both exchange field and spin-orbit coupling (SOC), so the spin-up and spin-down bands remain degenerate as shown in Fig.~\ref{Fig2}(b), and the zigzag $\alpha'$-BNR junction displays semiconducting behavior with a 0.1 eV bandgap. 
A spin gap proportional to the exchange field strength opens between the spin-up and spin-down bands as shown in Figs.~\ref{Fig2}(a) and \ref{Fig2}(c), which causes energy windows for perfect spin polarization (the regions indicated by blue and red dashed lines) are created. Here, fully spin-down-polarized states occur in the energy range -0.06 eV$<E<$ 0 eV, while fully spin-up-polarized states appear in the range -0.26 eV$<E<$-0.2 eV. Unlike an ideal metallic system, which allows the transmission of electrons with both spin orientations, the zigzag $\alpha'$-BNR junction selectively blocks one spin orientation. The transmitted spin orientation depends on energy of incoming electrons, which can be tuned via an external gate voltage. As a result, the zigzag $\alpha'$-BNR junction acts as a metal for one spin state and as an insulator for the opposite one, demonstrating its potential as a tunable half-metallic system. Moreover, the asymmetry between the spin-up and spin-down bands around the zero Fermi energy reflects the material's potential to produce a pure spin current in spin caloritronic applications.    

\begin{figure}[!ht]
\centering
\includegraphics[scale=0.55]{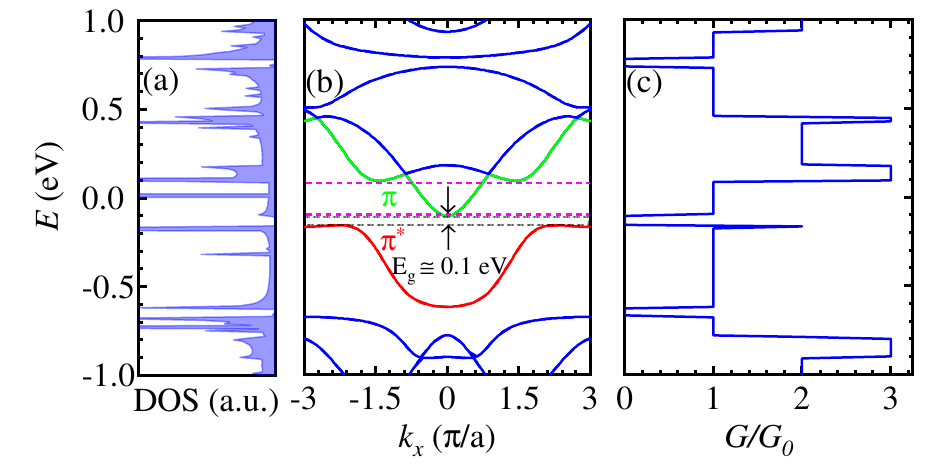}
\caption{(a) DOS, (b) band structure, (c) and conductance of the zigzag $\alpha'$-BNR. Distinct Van Hove singularities are evident in the DOS. The valence and conduction bands are indicated by red and green solid lines, respectively. The magnetization strength is set to \textit{$M_z$} = 0.098 eV.}
\label{Fig3}
\end{figure}

The density of states (DOS), band structure and quantum conductance of the $\alpha'$-BNR device are shown in Fig.~\ref{Fig3}. 
The sharp peaks in DOS indicate the Van Hove singularities in our one-dimensional system which are associated with the extremum points in the band structure (see Figs.~\ref{Fig3}(a) and~\ref{Fig3}(b)). Fig.~\ref{Fig3}(b) shows that the valence band maximum (VBM, red line) and the conduction band minimum (CBM, green line) near the Fermi energy are governed by the bonding $\pi$ and antibonding $\pi^*$ bands, which originate solely from the $p_z$ orbitals. A quasi-linear dispersion is observed near the $\pi$-band at $k_{x}a=\pm0.6\pi$, indicating a very small effective electron mass and, consequently, high electron mobility. Notably, the band dispersion in these regions closely resembles that of massless Dirac fermions.  
The semiconducting behavior of the $\alpha'$-BNR is also approved by the transmission, which is in good agreement with the band structure. Furthermore, the number of energy levels defines the transport channels at each energy level. As shown in Fig.~\ref{Fig3}(b), the number of transmission channels is equal to one in the energy range of -0.09 eV$<E<$ 0.08 eV (the region marked by the pink dashed line), whereas it becomes zero in the energy interval -0.15 eV$<E<$ -0.1 eV (the region indicated by the gray dashed line). 
\begin{figure}[t]
\centering
\includegraphics[scale=0.75]{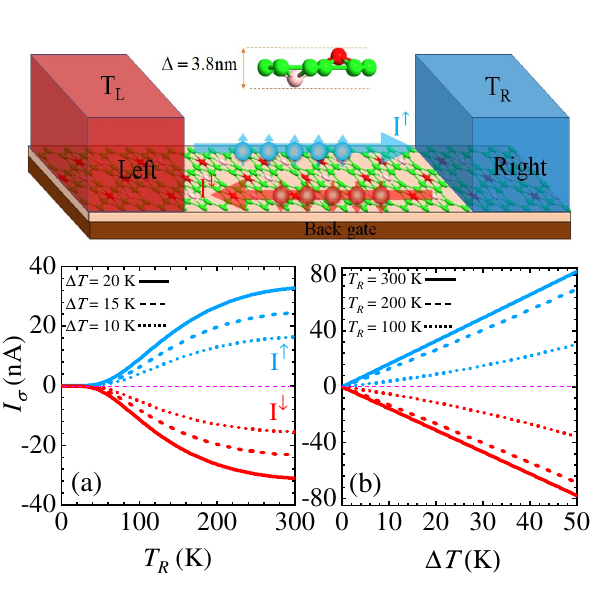}
\caption{(a) The spin-dependent currents $I_{\sigma}$ as a function of (a) right lead temperature $T_R$ for the different values of $\Delta{T}$ and (b) the temperature difference $\Delta{T}$ for ${T_R}=100, 200$ and 300 K.}
\label{Fig4}
\end{figure}
\begin{figure}[b]
\centering
\includegraphics[scale=0.37]{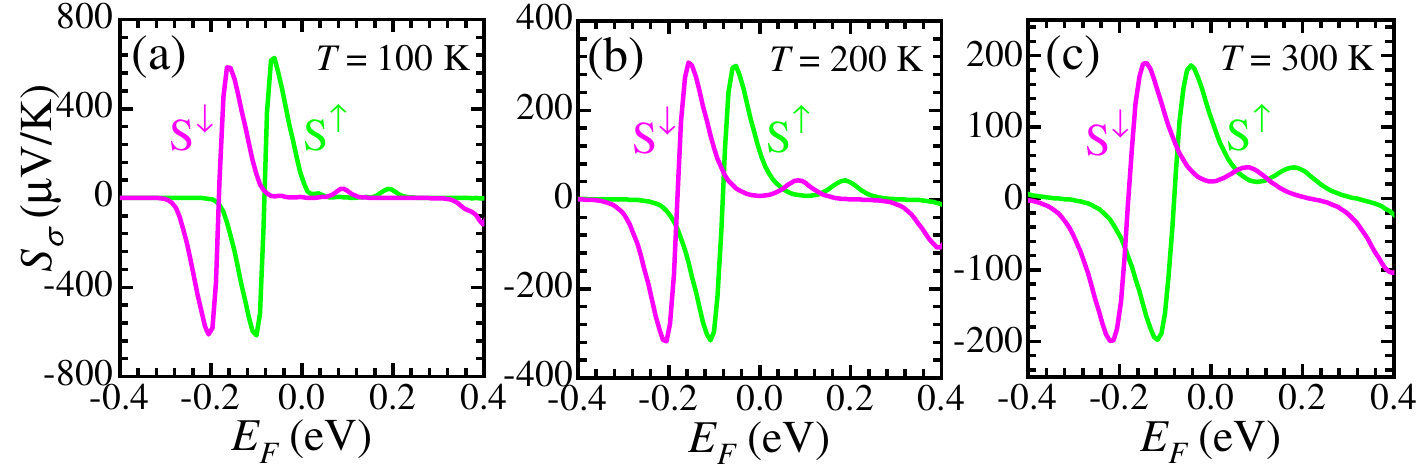}
\caption{(a) The spin-dependent Seebeck thermopower as a function of the Fermi energy for (a) \textit{T} = 100 K, (b) \textit{T} = 200 K, (c) \textit{T} = 300 K. The green and purple lines are for $S^{\uparrow}$ and $S^{\downarrow}$, respectively. Here, the magnetisation strength is \textit{$M_z$}=0.098 eV.}
\label{Fig5}
\end{figure}

Now, we examine thermal currents induced by a temperature difference (i.e., $\Delta{T}=T_{L}-T_{R}$) between the left and right leads. Using Eq. (21), the spin-dependent currents are plotted as functions of the right lead temperature $T_R$ for different values of $\Delta{T}$, as shown in Fig.~\ref{Fig4}(a). As observed in this figure, there is a threshold temperature ($T_{th}\approx{45}$ K for $\Delta{T}=10$ K ) below which the spin-dependent currents are zero ($T_{R}<T_{th}$). For $T_{R}>T_{th}$, the spin currents increase from zero (as $T_R$ increases), while $T_{th}$ shifts to lower values as the applied temperature difference increases. This behavior can be attributed to the broadening of the Fermi-Dirac distributions at larger $\Delta{T}$, which enhances the overlap between the available transmission channels of the junction. Furthermore, according to Eq. (21), the difference $f_{L}-f_{R}$ is an odd function with respect to the Fermi energy for $T_{L}>T_{R}$. Consequently, the temperature gradient breaks electron-hole symmetry, driving electrons and holes to flow in opposite directions. Because the spin-up (spin-down) current is primarily carried by holes (electrons), $I_{\uparrow}(I_{\downarrow})$ assumes positive (negative) values, as shown in Fig.~\ref{Fig4}. From $T_{R}=45$ K to $T_{R}=300$ K, the magnitudes of the thermally induced spin-up and spin-down currents are equal, resulting in a non-dissipative spin-dependent Seebeck effect in the zigzag $\alpha'$-BNR junction. In this regime, a pure thermal spin current satisfying 
$ I_{\uparrow}=-I_{\downarrow}$ is generated without an accompanying thermal charge current, i.e., $I_{c}=I_{\uparrow}+I_{\downarrow}=0$. This phenomenon provides an efficient mechanism for thermal spin injection, which is essential for spintronic applications. Accordingly, spin caloritronics is proposed as a promising approach for energy-harvesting devices that convert thermal energy into spin currents with minimal energy loss. Furthermore, as shown in Fig.~\ref{Fig4}(b), the thermal spin current increases linearly with increasing the temperature difference $\Delta{T}$.
    
The spin-dependent Seebeck thermopower $S^{\sigma}$ versus the Fermi energy at \textit{T} = 100, 200, and 300 K with the parallel configuration of magnetization is illustrated in Fig.~\ref{Fig5}. As shown, $S^{\sigma}$ changes sign with variations in the Fermi energy. According to Eqs. (23) and (24), the spin-dependent Seebeck thermopower comprises two primary contributions: one from holes ($E-E_{F}\leq{0}$) and the other from the electrons ($E-E_{F}\geq{0}$). Here, the positive and negative values of $S^{\sigma}$ originate from hole and electron transport, respectively, indicating that the spin-dependent channel is dominated by electrons (holes) in the negative (positive) regions of $S^{\sigma}$.This behavior drives spin-up and spin-down carriers in opposite directions, leading to the generation of a thermal spin voltage under a temperature gradient. Since the Seebeck coefficient $S^{\sigma}$ is directly related to the energy derivative of the spin-dependent transmission $T^{\sigma}$ or conductance $G^{\sigma}$, its maximum values occur near the edges of the spin-dependent band gap, where $G^{\sigma}$ varies rapidly. Furthermore, Fig.~\ref{Fig5} reveals neutral points at which the electron and hole contributions to the spin-dependent Seebeck thermopower cancel each other, satisfying $S^{\uparrow}+S^{\downarrow}=0$. At these points, although the thermal charge voltage vanishes (i.e., $V_c=0$), a finite thermal spin voltage is generated. Notably, the neutral point located at $E_F=-{0.13}$ eV remains unchanged with increasing temperature, indicating that no Fermi-level tuning is required to generate a pure spin current in the zigzag $\alpha'$-BNR junction. These results highlight the strong potential of the $\alpha'$-BNR junction for spin-caloritronic applications.

\begin{figure}[h]
\centering
\includegraphics[scale=0.50]{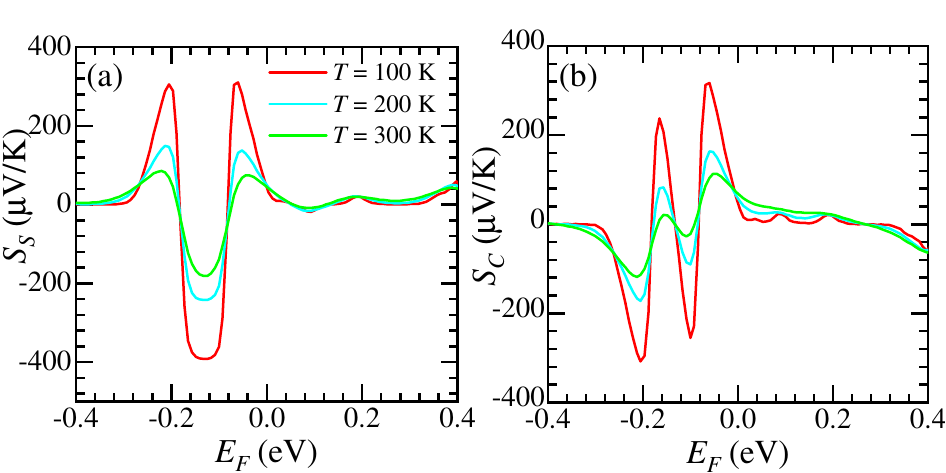}
\caption{(a) Spin thermopower and (b) charge thermopower as a function of the Fermi energy for different values of temperature. \textit{T} = 100 K, \textit{T} = 200 K, and \textit{T} = 300 K.}
\label{Fig6}
\end{figure}

\begin{figure}[t]
\centering
\includegraphics[scale=0.9]{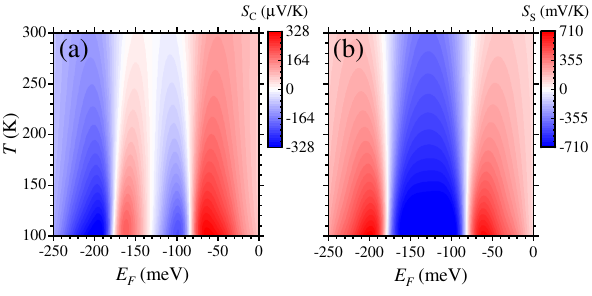}
\caption{Contour plots of (a) charge and (b) spin Seebeck coefficients as functions of the Fermi energy and temperature.}
\label{Fig7}
\end{figure}

\begin{figure}[t]
\centering
\includegraphics[scale=0.27]{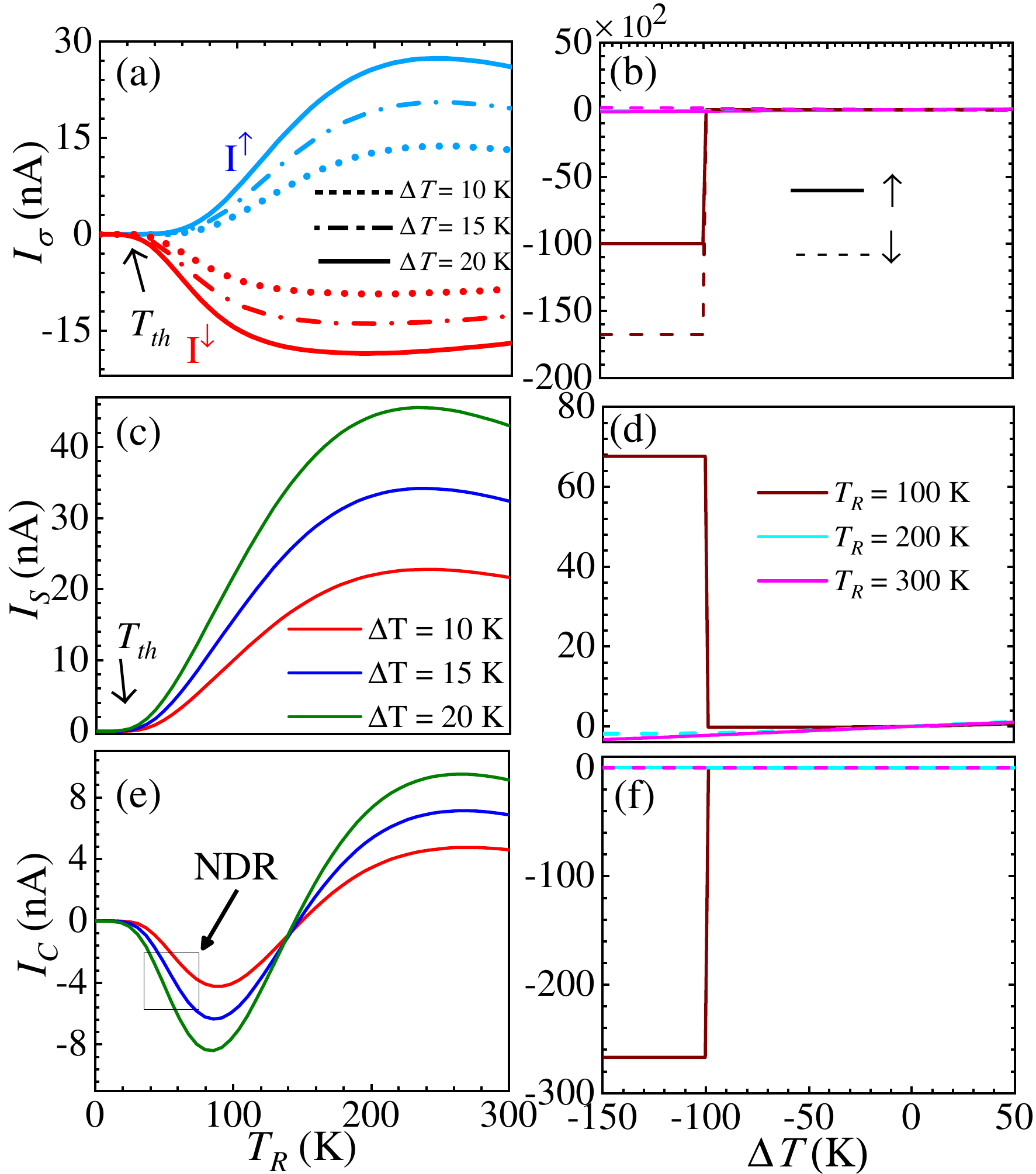}
\caption{(a) spin-dependent currents $I^{\uparrow}$ and $I^{\downarrow}$ as a function of the right lead temperature $T_R$ for different values of temperature difference $\Delta{T}=T_L-T_R$. (b) $I^{\uparrow}$ and $I^{\downarrow}$ versus $\Delta{T}$ for $T_R$ = 100, 200, and 300 K. (c) and (d) The pure spin current $I_S$ (= $I^{\uparrow}$ - $I^{\downarrow}$) versus $T_R$ and $\Delta{T}$,
respectively. (e) and (f) The charge electron current $I_C$ = ($I^{\uparrow}$ + $I^{\downarrow}$) versus $T_R$ and $\Delta{T}$, respectively.}
\label{Fig8} 
\end{figure}

The spin and charge Seebeck coefficients versus the Fermi energy for different temperatures are shown in Fig.~\ref{Fig6}. As can be seen, the spin-Seebeck coefficient decreases with increasing temperature. The spin Seebeck coefficient has a maximum value for different temperatures at the Fermi energy of -0.13 eV, where the charge Seebeck coefficient simultaneously vanishes. In addition, a net charge current can be driven solely by the applied temperature difference ($\Delta{T}$). At a specific Fermi energy (\textit{$E_F$} = -0.08 eV), a pure electronic current flows from the higher-temperature left lead to the lower-temperature right lead. Therefore, by appropriately tuning the Fermi energy, it is possible to selectively generate either a pure charge current or a pure spin current in the system. 
\\
Figure.~\ref{Fig7} presents contour plots of the charge and spin thermopowers as functions of Fermi energy and temperature. As shown in Fig.~\ref{Fig7}(a), the charge Seebeck coefficient vanishes about Fermi energy of -130 meV over a broad temperature range, while the spin Seebeck coefficient reaches its maximum at the same energy, as illustrated in Fig.~\ref{Fig7}(b). Moreover, both the charge and spin Seebeck coefficients decrease with increasing temperature and Fermi energy. These observations are consistent with the results discussed above.
To gain further insight, we plot the spin-dependent currents versus the right-lead temperature $T_R$ for various temperature differences $\Delta{T}$ = 10, 15, and 20 K ($\Delta{T} = T_L-T_R$) and the results are depicted in Fig.~\ref{Fig8}. As shown in Fig.~\ref{Fig8}(a), below the threshold temperature ($T_{th}$ = 42 K), both spin-up and spin-down currents are nearly zero. Applying a larger temperature difference enables the achievement of higher spin-dependent current as seen in Fig.~\ref{Fig8}(c). Moreover, the spin direction of the transmitted carriers can be switched from down to up by reversing the magnetization direction of the FM leads. As shown in Fig. 8(e), the charge current $I_c$ exhibits a peak at $T_R=80$ K and subsequently decreases. In fact, the zigzag $\alpha'$-BNR junction exhibits negative differential thermal resistance (NDTR). In low-temperature regime, the temperature gradient drives spin-dependent currents in opposite directions, with positive and negative polarities for spin-up and spin-down channels, respectively. Since the absolute magnitudes of the spin-up and spin-down currents are nearly equal, a pure spin-dependent Seebeck effect emerges. 

Figures~\ref{Fig8}(d) and \ref{Fig8}(f) present the spin and charge currents as functions of positive and negative temperature differences $\Delta{T}$ for various values of $T_R$. As shown, both currents exhibit a thermoelectric diode behavior at $T_R=100$, where nonzero spin and charge currents appear only under a negative temperature bias (i.e., $T_R>T_L$ or $\Delta{T}<0$). This effect can be understood from Eqs.~(21) and (25), which show that the thermal spin and charge currents depend on the spin-dependent transmission $T^{\sigma}(E)$, and the difference between the Fermi-Dirac distribution functions, $\Delta{f}=f_L-f_R$. At high temperatures, the slope of the Fermi-Dirac distribution function becomes very small (see, Eq.~(25)). Thus, when $T_{R}>T_{L}$ (i.e., $\Delta{T}<0$), $f_R$ varies slowly and is nearly uniform function with respect to $f_L$ around the Fermi energy. In this case, both $\Delta{f}$ and the transmission function $T^{\sigma}(E)$ are nonzero within the energy bands. Conversely, for $T_{L}>T_{R}$, $f_L$ is nearly uniform function with respect to $f_R$, and $\Delta{f}$ attains a large value in the band gap region where $T^{\sigma}(E)$ is zero. Consequently, the spin current $I_{\sigma}$ and the charge current $I_c$ vanish for $T_{L}>T_{R}$ or $\Delta{T}>0$. These results demonstrate that the zigzag $\alpha'$-BNR junction effectively operates as an ideal spin-Seebeck diode (SSD), making it highly suitable for spin-caloritronic applications.   
\begin{figure}[ht]
\centering
\includegraphics[scale=1.1]{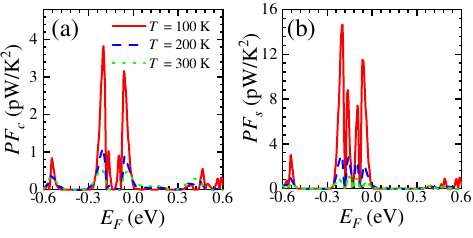}
\caption{(a) charge and (b) spin power factors as a function of the Fermi energy for different temperatures \textit{T} = 100, 200, and 300 K.}
\label{Fig9}
\end{figure}

Now, we study the power factor (PF) as a key attribute for assessing the efficacy of materials in thermoelectric device applications. The charge and spin PF as a function of the Fermi energy for different temperatures is represented in Fig.~\ref{Fig9}. The maximum value of the charge and spin PF decrease and widen by increasing the temperature. As we mentioned in Eq.~\ref{Eq.23}, the spin and charge Seebeck thermopowers have an inverse relation with the temperature, thus by increasing the \textit{T}, the spin and charge thermopowers reduce (see Fig.~\ref{Fig6}). This figure also clearly shows that the charge and spin PFs in the Fermi energy region -0.24 eV$<E_F<$0 become larger where the spin and charge thermopowers reache higher values (see Fig.~\ref{Fig6}). Moreover, the spin PF of $\alpha'$-BNR is nearly four times larger than the charge PF. This enhancement originates from the strong asymmetry in the spin-dependent electronic band structures, as illustrated in Fig.~\ref{Fig2}. Such a pronounced spin polarization around the Fermi level leads to a significant difference between the transport properties of spin-up and spin-down channels, thereby enhancing the spin thermoelectric response. Therefore, $\alpha'$-BNRs exhibit promising performance for applications in spin-based thermoelectric devices. 

\begin{figure}[ht]
\centering
\includegraphics[scale=1.1]{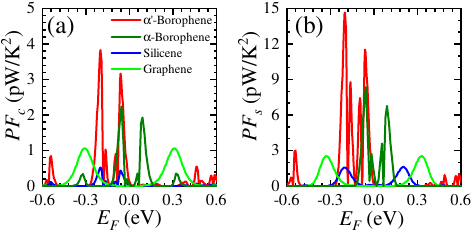}
\caption{(a) charge and (b) spin power factors as a function of the Fermi energy for different nanoribbon including $\alpha'$-borophene, graphene and silicene in the same conditions. The temperature is \textit{T} = 100 K.}
\label{Fig10} 
\end{figure}

Finally, to further clarify the high thermoelectric efficiency of $\alpha'$-BNR, we present a comparative analysis of the spin and charge PFs as a function of Fermi energy for BNRs in the $\alpha'$ and $\alpha$ phases, together with graphene and silicene nanoribbons and show the results in Fig.~\ref{Fig10}. The figure indicates that the $\alpha'$ phase has a significantly higher charge and spin PFs in comparison with the other 2D materials. This enhancement can be attributed to the unique electronic structure of the $\alpha'$ phase, which shows strong spin-dependent band asymmetry with high asymmetry between electrons and holes. In contrast, graphene nanoribbons possess relatively symmetric band structures and weaker spin polarization, resulting in lower spin and charge thermopowers. Similarly, silicene nanoribbons exhibit a reduced electron and hole asymmetry compared to $\alpha'$-BNR. The combination of high electronic conductivity and large spin-dependent Seebeck coefficient in $\alpha'$-BNR leads to an improved thermoelectric response. These findings highlight the superior potential of $\alpha'$-BNRs for next-generation spin thermoelectric applications.   
\section{SUMMARY}\label{sec4}
We have investigated the spin-dependent conductance of the $\alpha'$-BNR through the proximity of ferromagnetic material with the right and left leads. All the calculations have been performed using the NEGF approach inside the TB model framework. Our designed system shows half-metallicity and spin-filtering characteristics by applying an exchange magnetic field. We have also calculated the spin-dependent current using the Landauer-B\"uttiker formula. We find that by applying temperature differences and breaking the electron-hole symmetry in the system, the spin-up and spin-down currents flow in opposite directions which shows a pure spin current in the $\alpha'$-BNR. It can act as an ideal spin-Seebeck diode. Further, a negative differential spin-Seebeck effect through the compensation of thermal spin occurs in our proposed device. Finally, we have studied the temperature effect on the efficiency of the $\alpha'$-BNR known as the charge and spin power factors. It is shown in the vicinity of the Fermi level, by increasing the temperature the maximum magnitude of the charge and spin PFs decreases dramatically because the thermopower is reduced. Furthermore, by  comparing the the charge and spin PFs of $\alpha$ and $\alpha'$ phases of BNR, graphene and silicene nanoribbons, we have found the $\alpha'$-BNR has significantly higher value for charge and spin PFs. Our results show that $\alpha'$-BNR can be a highly efficient material in thermoelectric devices. 
\section*{CRediT authorship contribution statement}

\textbf{Farzaneh Ghasemzadeh:} Investigation, Data Curation, Visualization, Software. \textbf{Mohsen Farokhnezhad:} Conceptualization, Writing-original draft, Writing-review \& editing, Supervision, Validation, Methodology, Investigation. \textbf{Mahdi Esmaeilzadeh:} Writing-review \& editing, Project administration, Supervision. 
\vspace{1cm}
\section*{Declaration of Generative AI and AI-assisted technologies in the manuscript preparation process}

ChatGPT (OpenAI) was used solely to improve the language quality of the manuscript, including minor editing and paraphrasing. No AI tool was used for data analysis, interpretation, or generation of scientific content. All content was reviewed and approved by the authors, who take full responsibility for the manuscript.

\section*{Declaration of Competing Interest} 
Authors declare no competing interests.

\section*{Data availability}
Data will be made available on request.
\bibliographystyle{apsrev4-2}
\bibliography{ref}
\end{document}